\documentstyle[prl,aps,psfig,multicol]{revtex}

\newcommand{\anue}{\bar\nu_e}
\newcommand{\eps}{\varepsilon}
\newcommand{\mnue}{m_{\nu_e}}
\newcommand{\ncut}{N_{\rm cut}}

\begin{document}
\draft

\title{Electron Neutrino Mass Measurement by Supernova Neutrino Bursts
and Implications on Hot Dark Matter}

\author{Tomonori Totani}
\address{Department of Physics, School of Science,
The University of Tokyo,  \\
Tokyo 113, Japan \\
e-mail: totani@utaphp2.phys.s.u-tokyo.ac.jp}

\date{\today}

\maketitle

\begin{abstract}
We present a new strategy for measuring the electron
neutrino mass ($\mnue$) by future detection of a Galactic supernova in
large underground detectors such as the Super-Kamiokande (SK).
This method is nearly model-independent and one can get a mass
constraint in a straightforward way from experimental data
without specifying any model parameters for profiles of supernova neutrinos.
We have tested this method using virtual data generated from
a numerical model of supernova neutrino emission by
realistic Monte-Carlo simulations of the SK detection.
It is shown that this method is sensitive to $\mnue$ of $\sim$ 3 eV for
a Galactic supernova, and this range 
is as low as the prediction of the cold+hot dark
matter scenario with a nearly degenerate mass hierarchy of neutrinos,
which is consistent with the current observations of 
solar and atmospheric neutrino anomalies and 
density fluctuations in the universe.
\end{abstract}


\begin{multicols}{2}

\section{Introduction}
It is well known that detection of a neutrino burst from a
collapse-driven supernova by
large underground detectors gives us some constraints 
on neutrino masses, due to the delay of arrival times depending
on the neutrino energy as $\Delta t = 5.15 (D/10{\rm kpc})
(m_\nu/1{\rm eV})^2 (\eps_\nu / 10 {\rm MeV})^{-2}$ msec,
where $D$ is a distance to the supernova, $m_\nu$ the neutrino mass,
and $\eps_\nu$ the neutrino energy \cite{zatsepin}.
Some upper bounds on the electron neutrino
mass ($m_{\nu_e}$), ranging in 10--20 eV, 
have already been derived by a number of papers
using the historical data of SN1987A \cite{SN1987A}. 
On the other hand,
given the status of tritium $\beta$-decay experiments, experimental
upper limit on $m_{\nu_e}$ are also considered to be
$\sim$ 10--15 eV \cite{pdg}.
Therefore the next Galactic supernova expected in the near future and 
its detection by the currently emerging international network
of second-generation neutrino detectors, such as the Super-Kamiokande (SK)
\cite{SK} or SNO \cite{SNO}, would give an important opportunity 
of setting a more stringent constraint on $m_{\nu_e}$ than the current
astrophysical or experimental limits. Especially, the
normal water \v{C}erenkov detectors are the most sensitive to
electron antineutrinos ($\anue$ 's) and an enormous number
of $\anue p \rightarrow e^+ n$ events expected in the SK ($\sim$
5000--10000 events) for a supernova at the Galactic center ($D = 10$
kpc) would give much better statistics than that of SN1987A.

However, the electron neutrino mass 
measurement by supernova neutrinos generally suffers
significant uncertainties related to the original profiles of 
supernova neutrino emission, i.e., neutrino luminosity curve
and energy spectrum. The analyses on SN1987A data were based on the luminosity
decay during the cooling phase of hot neutron stars, during which
the majority of $\anue$'s is emitted. Because the decay time scale 
in this phase is $O$(10) seconds, it is difficult to
probe the arrival time delay shorter than this scale. This is why we could not
probe the mass scale smaller than
$\sim$ 10 eV for SN1987A ($D$ = 50 kpc).
Therefore it is clear that we have to devise a different strategy
which maximally utilizes the much larger number of expected events in the SK.
Although the cooling phase 
is $\sim$ 10 seconds long, the time scale of initial rise of neutrino
luminosity is much shorter: a recent numerical simulation of gravitational 
collapse and 
neutrino emission \cite{SN-sim} shows that this time scale for $\anue$'s
is 1--10 msec.
The prediction of this time scale by the current theory of
collapse-driven supernovae is robust because this is determined by
the time scale for the shock wave generated by the core bounce to
cross the neutrino sphere \cite{BKG}. This suggests that we can probe
the neutrino mass of $\sim$ 1 eV, at least in principle, by using
a sufficient number of events around the initial steep rise of neutrino
luminosity. In the following we propose a new strategy to set a
constraint on $\mnue$ from the initial rise of $\anue p$ events
assuming a detection by the SK. We then test this method by virtual
data of neutrino events detected by the SK, which are produced by
realistic Monte-Carlo simulations (MCs) with a numerical model of supernova
neutrino emission \cite{TSDW}. We find that $\mnue$ of $\sim$
3 eV can be probed by a future Galactic supernova.

\section{Getting Constraints on the Electron Neutrino Mass}
Strong $\anue$ emission suddenly breaks out when the shock wave passes
the neutrino sphere with a time scale of 1--10 msec, and after this
breakout the time variability of neutrino luminosity or energy spectrum
is on a scale of $\sim$ 1 second \cite{SN-sim,BKG,TSDW}. 
The signature of a finite neutrino mass which
we try to detect is the earlier arrival of high energy neutrinos
in the breakout.
Since events from the $\anue p \rightarrow e^+ n$ reaction are 
dominant, we treat all events as this reaction, and the validity of this
approximation will be checked later.
Suppose that we get a sequence of arrival time and detection 
energy of positrons as $(t_1, \eps_1)$, $(t_2, \eps_2)$, $\cdots$, 
$(t_N, \eps_N)$, where $N$ is the observed number of events, and
order of events is defined as $t_k < t_{k+1}$. 
Consider a transformation of detection time of events 
($t_k \rightarrow t_k'$)
for a given value of $\mnue$, which subtracts the
arrival time delay due to the assumed neutrino mass, as
\begin{equation}
t'_k = t_k - \frac{D \mnue^2}{2c (\eps_k + \Delta_{np})^2} \ ,
\end{equation}
where $\Delta_{np}$ is 
the neutron-proton mass difference (= 1.3 MeV).
Let $\eps_k'$ be the sequence of detection energy in 
increasing order of $t_k'$. Since the neutrino spectrum is roughly
constant after the breakout on a time scale of $\sim$ 1 second,
it is expected that 
the distribution of $\eps_k'$ is random without any correlation to
$t_k'$, if the assumed value of $\mnue$ is correct.
Here we define a measure of correlation between $t_k'$
and $\eps_k'$ in the first $\ncut$ events as follows:
\begin{equation}
S^2(\mnue) \equiv \sum_{k = 2}^{N_{\rm cut}} \frac{ \{ N_k(\eps_k') - (k-1) 
f(\eps_k') \}^2}{ (k-1) f(\eps_k') } \ ,
\end{equation}
where $N_k(\eps_k')$ is the number of events detected earlier than
$t_k'$ with energy greater than
$\eps_k'$, and $f(\eps)$ is the fraction of expected 
events with energy greater than $\eps$. 
We can calculate $f(\eps)$ from $\anue$ spectrum, the cross section of
$\anue p$ reaction, and detection efficiency of the SK.
If we use the Fermi-Dirac (FD)
distribution with zero chemical potential as the spectrum of 
neutrinos, we can straightforwardly calculate $S^2$ as a function of $\mnue$,
FD temperature $T_{\anue}$, and $N_{\rm cut}$
from a given experimental data set of $(t_k, \eps_k)$.
It is known that the real energy spectrum of supernova neutrinos is
slightly different from the pure black body radiation because of 
energy-dependent opacity of neutrinos. However we note that we are 
paying attention to how random 
the $\eps_k'$ distribution is, and a detailed shape of the
spectrum is not important in our analysis. This will be checked later.
Now let us consider the physical meaning of $S^2(\mnue, T_{\anue},
N_{\rm cut})$. If we approximately regard the Poisson distribution 
as the Gaussian distribution, the distribution of $S^2$ is
the $\chi^2$ statistics with $\ncut-1$ $(\sim \ncut)$ degrees of freedom, 
and hence it is expected that $S^2$ obeys the $\chi^2$ distribution
if the assumed $\mnue$ is correct. On the other hand, if the assumed
$\mnue$ is significantly different from the true value, only high-
or low-energy neutrinos will arrive earlier and $S^2$ will become
larger than the expectation from the $\chi^2$ distribution. 
Similarly, incorrect values of
$T_{\anue}$ will lead to unexpectedly large $S^2$.
Therefore $S^2$ is expected to take
the minimum at the correct values of $\mnue$ and $T_{\anue}$. Hereafter
we always take a value of $T_{\anue}$ which minimizes $S^2$.
Then we get a constraint on $\mnue$ with $n$ sigma confidence level,
for a given value of $\ncut$ as
\begin{equation}
S^2(\mnue, N_{\rm cut}) < \min_{\mnue} S^2(\mnue, \ncut) + 
n \sqrt{2 \ncut},
\end{equation}
with the best-fit value of $\mnue^{\rm fit}(\ncut)$ which minimizes
$S^2$, where $2 \ncut$ is the variance of the 
$\chi^2$ distribution with $\ncut$ degrees of freedom.
We stress that this strategy does not require any
specification of model parameters, and constraints on $\mnue$ can
be calculated in a straightforward way from experimental data.
It should also be noted that we have implicitly assumed in the above argument
that the distance to a supernova is known, although in a future
detection it may be unknown.
In such case the above strategy is still applicable but
we can only get constraint on $D \mnue^2$.

\section{Test for the Strategy by Monte-Carlo Simulations}
In the following of this letter, we give a test about the reliability of this 
strategy, by using neutrino emission profiles of 
a one-dimensional numerical model of supernova explosion which does not assume
any particular energy distribution of neutrinos \cite{SN-sim}. 
This is a model of SN1987A with a 
main-sequence mass of $\sim 20 M_\odot$. We have made virtual data sets
of $(t_k, \eps_k)$ by Monte-Carlo simulations of the SK detector
supposing a supernova at a distance of $D$ = 10 kpc. The
MC simulation is described in Ref. \cite{TSDW} which takes
account of the SK detection efficiency, energy resolution, and
reaction modes of $\anue p$ absorption, $\nu e$ scatterings, and
charged-current
$\nu_e (\anue)$ absorptions into oxygen. Therefore we can check 
the validity of the approximations in the proposed method, i.e., assuming
the FD distribution and regarding all events as $\anue p$ events.
We have made four MC realizations with simulated neutrino masses
($\mnue^{\rm MC}$) of 0, 3, 5, and 7 eV,
and applied the method to these data
varying the value of $\ncut$. We have used events with detection energy
greater than 10 MeV, to avoid the background noise of $(\nu, \nu'
p \gamma)$ and $(\nu, \nu' n \gamma)$ reactions on $^{16}$O \cite{LVK96}.
The obtained best-fit $\mnue^{\rm fit}$ and 2 sigma (95\%
C.L.) lower- and upper-limits ($\mnue^l$ and $\mnue^u$)
are shown in Fig. 1 as functions of $\ncut$.
If the strategy correctly detects the signature 
of a finite neutrino mass, the best-fit $\mnue^{\rm fit}$ should 
not vary with $\ncut$. In other words, the constancy of $\mnue^{\rm fit}$ 
against $\ncut$
gives an important consistency check of this analysis.
The figure shows that, for $\mnue^{\rm MC}$ = 3, 5, and 7 eV,
$\mnue^{\rm fit}$ is 
almost constant at the simulated values
in $\ncut \sim$ 100-300, suggesting that
this method correctly detects a finite neutrino mass
if we use neutrinos of the first 200-300 events,
i.e., during 60--80 msec after the core bounce. 
One can see a slight systematic decrease of
$\mnue^{\rm fit}$ in $\ncut \agt$
300, and this is an effect of spectral hardening of $\anue$'s, 
which begins from $\sim$ 100 msec after the core bounce
as a signature of the delayed explosion mechanism 
\cite{BKG,TSDW}. It should be noted that, in the prompt explosion 
scenario, stellar envelope is expelled in a very short time scale
of 1--10 msec which is the same as that of the shock breakout, 
and neutrino spectrum is rather constant after the breakout.
Therefore the proposed strategy would work even better in prompt explosions.
The optimal value of $\ncut$ depends on the distance
and profiles of neutrino emission, and it should be determined by
the constancy in the  $\mnue^{\rm fit}$-$\ncut$
diagram produced from real data detected in the future.

We next estimate the sensitivity of the proposed strategy 
by statistical average of many MC realizations. One hundred MC
realizations are generated from the supernova model with $D$ = 10 kpc
for each of three values of $\mnue^{\rm MC}$ = 0, 3, and 5 eV, and 
the proposed method
is applied to these data with $\ncut$ = 200. The average of 
$\mnue^{\rm fit}, \mnue^l,$ and $\mnue^u$ for the 100 MCs are shown
in Table 1 with 1 $\sigma$ statistical fluctuations. In order to check
the validity of estimated confidence levels, the probability that
this method gives incorrect results (i.e., $\mnue^l > \mnue^{\rm MC}$ or
$\mnue^u < \mnue^{\rm MC}$) 
is also shown in this table, as well as the probability
of detecting finite $\mnue$ (i.e., $\mnue^l > 0$).
These results suggest that the estimated
confidence levels are roughly valid for $\mnue^{\rm MC}$ = 0 and 3 eV. 
For $\mnue^{\rm MC}$ = 5
eV, $\mnue^{\rm fit}$ is systematically smaller than $\mnue^{\rm MC}$
and 28 trials out of 100 MCs give 
incorrect results of $m_u < 5$ eV. However, this systematic
error is not greater than 1 eV and 
we can detect finite $\mnue$ with a probability of 99 \% if 
$\mnue$ = 5 eV. The origin of this systematic error is difficult
to understand clearly, but probably it is the gradual hardening of the
neutrino spectrum. The probability of detecting finite mass
is about 50\% for $\mnue$ = 3 eV, and we can conclude that 
this method is marginally sensitive to the electron neutrino mass of 3 eV,
and can easily detect $\mnue$ of 5 eV. This strategy also gives a
fit of neutrino effective temperature $T_{\anue}^{\rm fit}$, and
the average of $T_{\anue}^{\rm fit}$ is also given in the table.
This $T_{\anue}^{\rm fit}$ agrees well with the true neutrino spectrum,
considering that average $\anue$ energy is 3.15$T_{\anue}$ in FD
distribution and that of the numerical supernova model in this early phase
is about 10--12 MeV. (FD-fit average energy is a little lower
than the true value because of the deviation of the true spectrum from
the FD distribution. See Fig. 9 of ref. \cite{TSDW}.)

Now let us consider the dependence of the proposed strategy on the distance
to a supernova. There are two competing effects: available number of 
events becomes smaller with increasing distance, while the time
delay due to the finite mass increases. In order to see which is more
effective, we have tested the proposed method against supernovae
at $D$ = 5 and 20 kpc, with $\mnue^{\rm MC}$ = 0 eV. 
Statistical average of $\mnue^u$ for
100 MC simulations is
2.4 $\pm$ 0.6 and 3.1 $\pm$ 0.7 [eV] for $D$ = 5 and 20 kpc cases,
respectively (Table 1). Here we have used 400 and 100 as 
the values of $\ncut$, which are found to be appropriate 
from $\mnue^{\rm fit}$-$\ncut$ 
diagrams. Combined with the fact that the average of $\mnue^u$
for $D$ = 10 kpc case is 2.8 $\pm$ 0.7 [eV],
the sensitivity becomes slightly better with 
decreasing distance, but the dependence is very weak and smaller than
statistical dispersion. Therefore we conclude that 
the sensitivity of the proposed
method is roughly the same for any collapse-driven supernova in our Galaxy.
In the above estimate, we have abandoned low-energy events below
10 MeV, to avoid $\gamma$-ray events induced by neutral current
reactions of $\nu_\mu$ (or $\nu_\tau$) with $^{16}$O which are expected
to be roughly the same number with $\anue p$ events 
in 5--10 MeV \cite{LVK96}.
If we could somehow remove or effectively subtract
these noises and apply the method to
all events with detection energy greater than
5 MeV (threshold of the SK), then
the average of $\mnue^u$ could be as low as 1.6 $\pm$ 0.4 [eV]
for a supernova at $D$ = 10 kpc (Table 1). 

\section{Discussion}
Finally we discuss some implications of the reported sensitivity of
supernova neutrinos to the electron neutrino mass. Currently there are some
hints on nonzero neutrino masses in astrophysical and cosmological
observations, and here we consider  the following three: 
the solar and atmospheric neutrino anomalies and hot
dark matter in the universe. The standard cold dark matter model of
structure formation normalized by the COBE data with $\Omega_0 = 1$ 
is known to be inconsistent with the clustering properties of
galaxies or clusters of galaxies, and the cold+hot dark matter model 
with $\Omega_{\nu} \sim$ 0.2--0.3 
is one of some possibilities to resolve this discrepancy,
where $\Omega_{\nu}$ is the fraction of
hot dark matter in the critical density of the universe (e.g.,
Ref. \cite{primack}).
The most promising solution for the solar neutrino problem
is the MSW solutions with $\Delta m^2 \sim 10^{-5}$ [eV$^2$]
\cite{solar}, and  the atmospheric neutrino anomaly can be explained by
the neutrino oscillation with $\Delta m^2 \sim 10^{-3}$--$10^{-1}$
[eV$^2$] \cite{atmos}. The only way to combine these hints without
sterile neutrinos is an almost degenerate mass hierarchy of neutrinos 
\cite{ADM}, with $
m_\nu = 4.6 (h/0.7)^2 (\Omega_{\nu}/0.3)$ [eV]
for all the three generations of neutrinos, where $h$ is the Hubble 
constant $H_0$/(100km/s/Mpc).
Oscillations between $\nu_e \leftrightarrow \nu_\mu$
and $\nu_\mu \leftrightarrow \nu_\tau$ give the solutions for
the solar and atmospheric neutrino problems, respectively.
The proposed method to constrain $\mnue$ from supernova neutrino bursts
can probe $\mnue$ as low as this scale and might detect a finite
$\mnue$ if the mass hierarchy of neutrinos was actually degenerate.

Let us briefly discuss effects of possible neutrino oscillations.
If the mixing angle between $\nu_e$ and others is order unity,
the observed $\anue$ spectrum would be a mixture of original $\anue$
and $\bar\nu_\mu$ (or $\bar\nu_\tau$) due to the vacuum oscillation. 
(The MSW matter oscillation in supernovae is not effective for antineutrinos
unless the mass hierarchy is inverse.)
If the hierarchy of
neutrino masses is not degenerate, i.e., $m_1 \ll m_2 \ll m_3$
as observed in charged leptons, the mixture of light 
neutrinos with negligible time delay and heavy neutrinos with 
significant delay would make the proposed method inapplicable
to measurement of $\mnue$. However, in case of the almost degenerate
hierarchy, the effect of oscillation is only deformation of 
observed $\anue$ spectrum due to contamination of original $\bar\nu_\mu$'s,
and the proposed method is still applicable
because a detailed shape of $\anue$ spectrum is not important
in this method.


The author would like to thank J. R. Wilson and H. E. Dalhed
for providing the output of their simulation of a supernova explosion.
He is also grateful to an anonymous referee for useful comments.
This work has been supported by the Grant-in-Aid for the 
Scientific Research Fund No.3730 of the Ministry of Education, Science,
and Culture in Japan.

\begin{figure}
  \begin{center}
    \leavevmode\psfig{figure=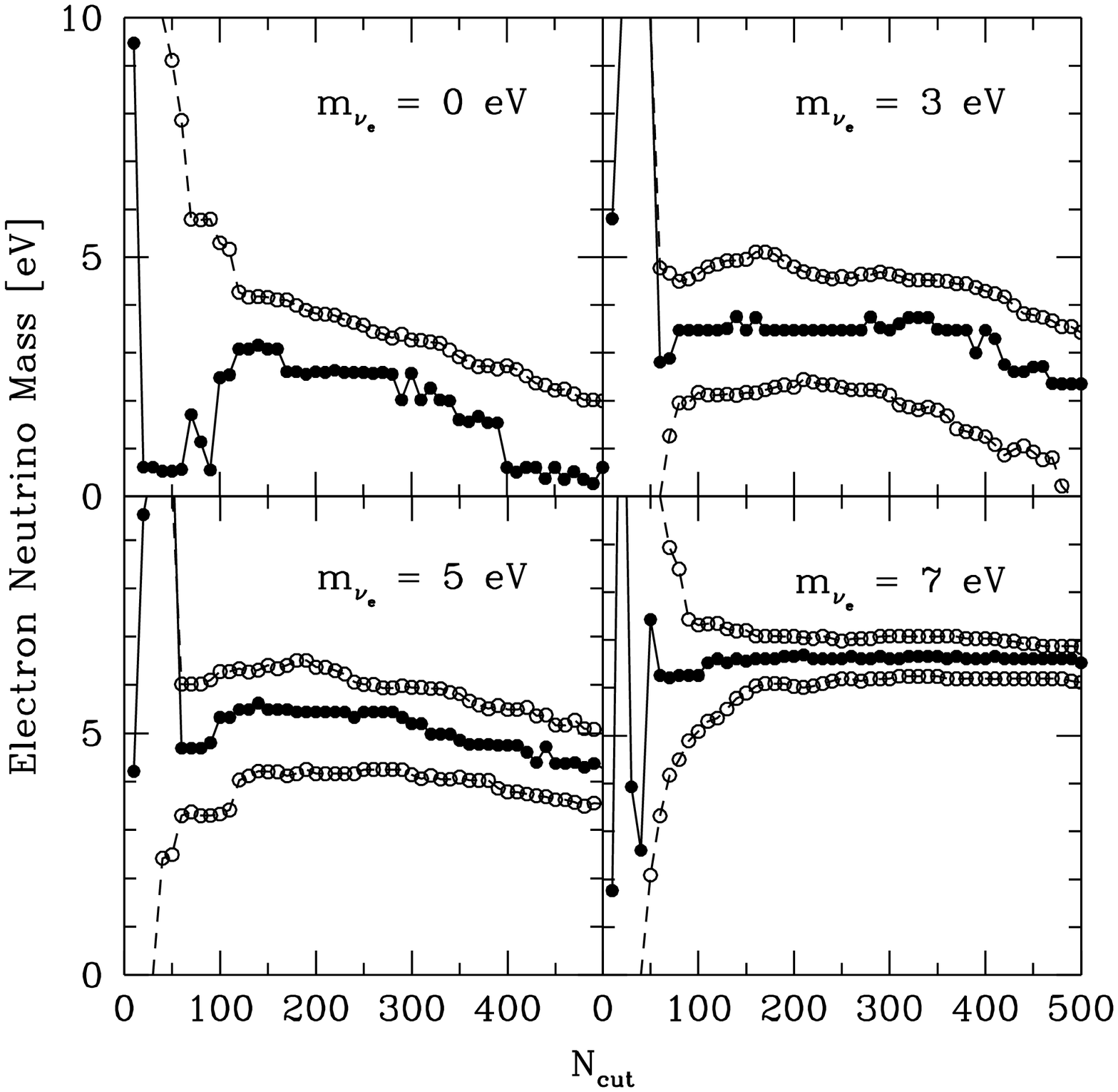,width=7.8cm}
  \end{center}
\end{figure}
{\footnotesize Figure 1.
The result of the proposed strategy of measuring $\mnue$
applied to four Monte-Carlo realizations of the SK detection of
a supernova at the Galactic center
with simulated $\mnue^{\rm MC}$ of 0, 3, 5, and 7 eV.
Best fit $\mnue$ (filled circles)
and 95 \% C.L. lower- and upper-limits (open circles) are shown as
functions of $\ncut$.
}

\end{multicols}

\begin{table}
  \begin{tabular}{cccccccccccc}
    &&&& \multicolumn{4}{c}{Average of 100 MCs} & &
    \multicolumn{3}{c}{Probability [\%]} \\
    \cline {5-8} \cline{10-12}
    $D$ [kpc] & $\mnue^{\rm MC}$ [eV] & $\ncut$ & $\eps_{\rm th}$ [MeV]
    & $\mnue^l$ [eV] & $\mnue^{\rm fit}$ [eV] &
    $\mnue^u$ [eV] & $T_{\anue}^{\rm fit}$[MeV]
    && $\mnue^l > 0$ & $\mnue^l > \mnue^{\rm MC}$ &
    $\mnue^u < \mnue^{\rm MC}$ \\
    \hline
    10 & 0 & 200 & 10 & 0.0$\pm$0.0 & 0.81$\pm$0.78 & 2.8$\pm$0.7 
    & 2.7$\pm$0.1 && 0 & 0 & 0 \\
    10 & 3 & 200 & 10 & 0.77$\pm$0.95 & 2.7$\pm$0.9 & 3.9$\pm$0.6
    & 2.8$\pm$0.2 && 47 & 3 & 5 \\
    10 & 5 & 200 & 10 & 3.2$\pm$0.7 & 4.4$\pm$0.5 & 5.3$\pm$0.5 
    & 2.8$\pm$0.2 && 99 & 1 & 28 \\
    5 & 0 & 400 & 10 & 0.04$\pm$0.22 & 0.74$\pm$0.85 & 2.4$\pm$0.6 
    & 2.6$\pm$0.1 && 5 & 5 & 0 \\
    20 & 0 & 100 & 10 & 0.05$\pm$0.05 & 0.92$\pm$0.91 & 3.1$\pm$0.7 
    & 2.7$\pm$0.2 && 1 & 1 & 0 \\
    10 & 0 & 200 & 5 & 0.0$\pm$0.0 & 0.35$\pm$0.40 & 1.5$\pm$0.4
    & 2.8$\pm$0.1 && 0 & 0 & 0 \\
  \end{tabular}
  \caption{The results of application of the proposed method for measuring
    $\mnue$ to virtual data of supernova neutrinos produced by
    100 Monte-Carlo simulations of the SK detection, where $D$ is the distance
    to a supernova, $\ncut$ the number of events used, $\eps_{\rm th}$
    the threshold energy in the analysis, and $\mnue^{\rm fit}$ and
    $\mnue^{l}$ ($\mnue^{u}$) are the best-fit mass and
    95 \% C.L. lower (upper) limits, respectively.
}
\end{table}


\begin{references}
\bibitem{zatsepin}
G. I. Zatsepin, Zh. Eksp. Teor. Fiz. Pis'ma {\bf 8} (1968) 333 [JETP Lett. 8
(1968) 205]

\bibitem{SN1987A}
T. J. Loredo and D. Q. Lamb, in {\it Fourteenth Texas Symposium on
Relativistic Astrophysics}, edited by E.J. Fenyves,
Ann. N.Y. Acad. Sci {\bf 571}, 601 (1989),
and references therein.

\bibitem{pdg}
Particle Data Group, \prd {\bf 54}, 1 (1996).

\bibitem{SK}
Y. Totsuka, Rep. Prog. Phys.  {\bf 55}, 377 (1992);
K. Nakamura, T. Kajita, M. Nakahata, and A. Suzuki, 
in {\it Physics and Astrophysics of Neutrinos},
edited by M. Fukugita and A. Suzuki
 (Springer-Verlag, Tokyo, 1994), p249.

\bibitem{SNO}
G. T. Ewan, in Proceedings of the Supernova Watch Workshop, Santa
Monica, California, 1990 (unpublished).



\bibitem{SN-sim}
The profiles of neutrino emission of a numerical supernova model used in this 
paper are described in Ref. \cite{TSDW}, and see also R. Mayle, 
Ph D. Thesis, Univ. of California (1985); J. R. Wilson, R. Mayle,
S. Woosely, and T. Weaver, Ann. NY Acad. Sci, {\bf 470}, 267 (1986);
R. Mayle, J. R. Wilson, and D. N. Schramm, \apj, {\bf 318}, 288 (1987).

\bibitem{BKG}
A. Burrows, D. Klein, and R. Gandhi, \prd {\bf 45}, 3361 (1992).

\bibitem{TSDW}
T. Totani, K. Sato, H. E. Dalhed, and J. R. Wilson, to appear in
Astrophys. J. (1997) (astro-ph/9710203). 

\bibitem{LVK96}
K. Langanke, P. Vogel, and E. Kolbe, \prl, {\bf 76}, 2629 (1996).

\bibitem{primack}
J. R. Primack, in {\it Critical Dialogues in Cosmology}, 
edited by N. Turok (World Scientific, 1996) (astro-ph/9610078).

\bibitem{solar}
A. Yu. Smirnov, in {\it Neutrino '96}, edited by
K. Enqvist, K. Huitu, and J. Maalampi (World Scientific, 1997), p38.

\bibitem{atmos}
T. K. Gaisser, in {\it Neutrino '96}, edited by
K. Enqvist, K. Huitu, and J. Maalampi (World Scientific, 1997), p211.

\bibitem{ADM}
D. Caldwell and R. N. Mohapatra, \prd {\bf 48}, 3259 (1993);
{\bf 50}, 3477 (1994); S. T. Petcov and A. Yu. Smirnov, Phys. Lett. B
{\bf 322}, 109 (1994); A. S. Joshipura, Z. Phys. C{\bf 64}, 31 (1994).

\end{references}
\end{document}